\documentclass{moriond}

\def\be{\begin{equation}}
\def\ee{\end{equation}}
\def\bea{\begin{eqnarray}}
\def\eea{\end{eqnarray}}

\newcommand{\Photo}{\includegraphics[height=35mm]{mypicture}}

\usepackage[version=3]{mhchem}

\newcommand{\bb}{$0\nu \beta \beta$} 
\newcommand{\bbvv}{$2\nu \beta \beta$} 

\newcommand{\mbb}{m_{\beta \beta}} 

\newcommand{\el}{\text{e}}

\newcommand{\ckky}{counts\,$\keV^{-1}\,\kg^{-1}\,\yr^{-1}$}

		\newcommand{\mK}{\text{mK}}

\newcommand{\yr}{\text{yr}}			\newcommand{\kg}{\text{kg}}

		\newcommand{\keV}{\text{keV}}

\begin{document}

	\vspace*{4cm}
	\title{Results of CUORE}
	\maketitle

	{\noindent \centering
S.~Dell'Oro$^{1}$, D.~Q.~Adams$^{2}$, C.~Alduino$^{2}$, K.~Alfonso$^{3}$, F.~T.~Avignone~III$^{2}$, O.~Azzolini$^{4}$, \\
G.~Bari$^{5}$, F.~Bellini$^{6,7}$, G.~Benato$^{8}$, M.~Biassoni$^{9}$
% ...
\\
}

\vspace{4mm}

\noindent
$^{1}$ Center for Neutrino Physics, Virginia Polytechnic Institute and State University, Blacksburg, Virginia 24061, USA \\
$^{2}$ Department of Physics and Astronomy, University of South Carolina, Columbia, SC 29208,~USA \\
$^{3}$ Department of Physics and Astronomy, University of California, Los Angeles, CA 90095, USA \\
$^{4}$ INFN -- Laboratori Nazionali di Legnaro, Legnaro (Padova) I-35020, Italy \\
$^{5}$ INFN -- Sezione di Bologna, Bologna I-40127, Italy \\
$^{6}$ Dipartimento di Fisica, Sapienza Universit\`{a} di Roma, Roma I-00185, Italy \\
$^{7}$ INFN -- Sezione di Roma, Roma I-00185, Italy \\
$^{8}$ Department of Physics, University of California, Berkeley, CA 94720, USA \\
$^{9}$ INFN -- Sezione di Milano Bicocca, Milano I-20126, Italy \\
% ...

\abstracts{
	The Cryogenic Underground Observatory for Rare Events (CUORE) at the Laboratori Nazionali del Gran Sasso, Italy, is the world's largest bolometric experiment.
	The detector consists of an array of 988 \ce{TeO_2} crystals, for a total mass of 742\,kg.
	CUORE is presently in data taking, searching for the neutrinoless double beta decay of \ce{^{130}Te}.
	CUORE is operational since the spring of 2017. The initial science run already allowed to provide the most stringent limit
	on the neutrinoless double beta decay half-life of \ce{^{130}Te}, 
	and to perform the most precise measurement of the two-neutrino double beta decay half-life. 
	Up to date, we have more than doubled the collected exposure.
	In this talk, we presenteded the most recent results and discuss the present status of the CUORE experiment.
}

\section{Introduction}

	Neutrinoless double beta decay (\bb,~\cite{Furry:1939qr}) is a rare nuclear process not predicted by the Standard Model in which a pair of neutrons inside a nucleus transforms into a 
	pair of protons, with the emission of two electrons: $(A,Z) \to (A,Z+2) + 2\el^-$.
	This transition clearly violates the conservation of the number of leptons. The observation of \bb~would thus demonstrate that the lepton number is not a symmetry of nature.
	At the same time, \bb~provides a key tool to study neutrinos by probing whether their nature is that of Majorana particles and providing us with important information on the 
	neutrino absolute mass scale and ordering~\cite{Dell'Oro:2016dbc}.

	The huge impact on Particle Physics has motivated and continues to motivate a strong experimental effort to search for \bb.
	Among the experiments searching for \bb, CUORE~\cite{Alduino:2017ehq}, acronym for Cryogenic Underground Observatory for Rare Events, is looking for the transition: 
	\ce{^{130}Te \to \ce{^130}Xe + 2e^-}.

	CUORE is located at the Laboratori Nazionali del Gran Sasso, Italy ($\sim 3600$\,m\,w.\,e.) and is presently in data-taking. The experiment is expected to
	collect data for a total of five years of live-time.
	
\section{CUORE detector}

	The CUORE detector comprises an array of 988 $5\times 5\times 5$\,cm$^3$ \ce{^{nat}TeO_2} crystals arranged into 19 towers of 13 4-crystal floors~\cite{Alduino:2016vjd}. 
	Each crystal has a mass of 750\,g, giving a total detector mass of 742\,kg, i.\,e.\ 206\,kg of \ce{^{130}Te}.
	The crystals are operated as cryogenic bolometers.

	Bolometers are calorimeters in which the energy released inside an absorber by an interacting particle is converted into phonons and measured via temperature variation. 
	These detectors can only be operated at cryogenic temperatures in order to minimize the heat capacity, since the intrinsic response of a calorimeter is proportional to this parameter.
	In the case of CUORE, the working temperature is about 10\,mK, where the heat capacity of the \ce{TeO_2} crystals is $\sim 100\,\mu$K per MeV.
	To detect any slight variation in temperature, each CUORE crystal is instrumented with a neutron transmutation doped (NTD) \ce{Ge} thermistor.
	Furthermore, all the crystals are also instrumented with a \ce{Si} heater, to stabilize the detector response by cyclically delivering a fixed (and extremely precise) amount of energy
	to the bolometers.
	 
	The detector assembly took almost two years, from September 2012 to July 2014. 
	Thanks to specifically designed procedures~\cite{Buccheri:2014bma}, the CUORE crystals were never exposed to air (thus avoiding the risk of contamination by \ce{Rn}) 
	from the moment of the polishing after growth until the installation of the detector.
	This latter operation was performed in summer 2016, after the completion of the commissioning of the cryogenic system.
	In the meanwhile, for about two years, the towers were stored inside the CUORE clean room into sealed containers constantly flushed with clean \ce{N_2}.
	
\section{CUORE cryostat}

	\begin{figure}[t]
		\centering
		\includegraphics[width=.4\columnwidth]{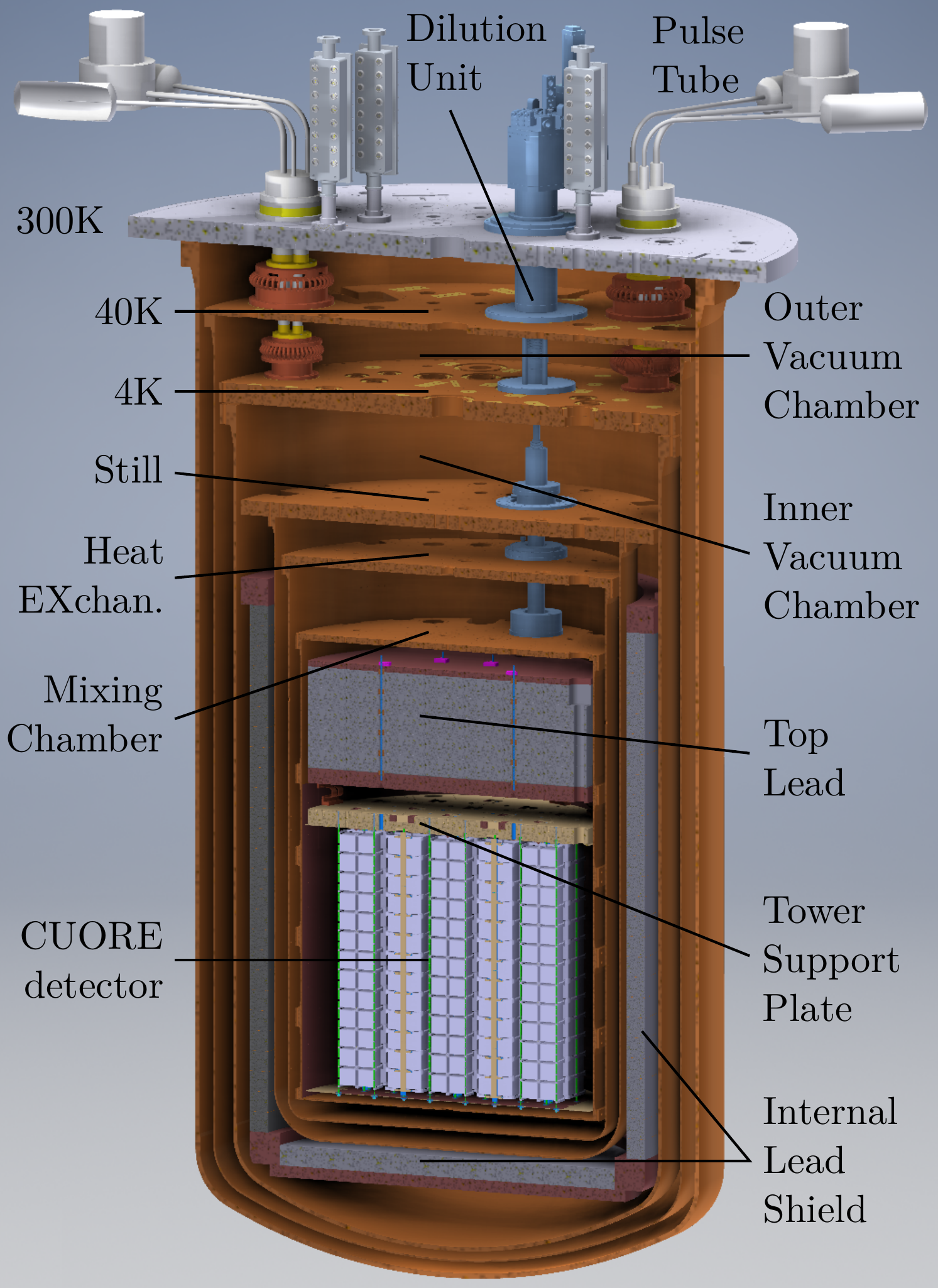}
		\caption{Rendering of the CUORE cryostat. The different thermal stages, vacuum chambers, cooling elements and lead shields are indicated.}
		\label{fig:cryostat_schematic}
	\end{figure}

	Given the huge size and mass, the CUORE detector could not be housed in any standard cryostat.
	In order to operate the detector, a custom cryogenic system had thus to be designed and constructed, satisfying very stringent experimental requirements in terms of high cooling power, 
	low noise environment and low radioactivity content (Fig.~\ref{fig:cryostat_schematic}, \cite{Alduino:2019xia}).

	The CUORE cryostat is a large custom cryogen-free cryostat cooled by 5 Pulse Tube Refrigerators and by a high-power \ce{^3He}/\ce{^4He} Dilution Unit with $3\,\mu$W at 10\,mK.
	The cryostat comprises six nested high-purity-copper vessels, the innermost of which encloses an experimental volume of about $1$\,m$^3$.
	The various stages thermalize to different temperatures, from room temperature to $\sim 10\,\mK$, and are identified by their approximate temperatures: 
	300\,K, 40\,K, 4\,K, 800\,mK or Still, 50\,mK or Heat EXchanger (HEX), and 10\,mK or Mixing Chamber (MC).
	At the center, the Tower Support Plate (TSP) holding the detector is attached to a dedicated suspension system in order to reduce the amount of vibrations.
	The 300\,K and the 4\,K vessels are vacuum-tight and define two vacuum volumes called the Outer Vacuum Chamber (OVC) and the Inner Vacuum Chamber (IVC).
	The detector is shielded from the external radioactivity by two lead shields placed inside the IVC.
	The Inner Lead Shield (ILS) stands between the 4\,K and the Still stages and provides side shielding and shielding from below. 
	This shield is made of ancient Roman lead, with extremely low concentration of \ce{^{210}Pb} ($<4\,$mBq\,kg$^-1$~\cite{roman_lead:1998}).
	The Top Lead is positioned below the MC plate and provides shielding from above.
	The whole cryostat is protected from the environmental radioactivity by the external shield made of 70\,t of lead and borated polyethylene.
	
	To cooldown the detector to its working temperature, almost one month is required in order to extract more than $7\cdot 10^8\,$J of enthalpy from the system. 
	The initial phase of the cooldown process is driven by a dedicated Fast Cooling System, that circulates \ce{He} gas through an external cooling circuit 
	and injects it directly into the IVC. Then, the Pulse Tubes bring the inner cryostat stages down to about 4\,K and the Dilution Unit completes the cooldown 
	of the Still, HEX and MC stages (including the detector).

	The cooldown of CUORE took place between December 2016 and January 2017.
	Indeed, after the cryostat construction, a period of about four years was required for the commissioning of the cryogenic system, before the installation of the CUORE detector.
	The commissioning was long and complex. It involved several test and cooldowns to integrate the numerous custom components and to check the system performance.
	Nonetheless, at the end of this process, the success of the CUORE cryostat marked a major milestone in the history of low-temperature detector techniques and opened 
	the way for large bolometric arrays (tonne-scale) for rare event physics.
	
\section{Initial results from CUORE}

	The initial few months of operation of CUORE were devoted to the detector characterization and optimization, i.\,e.\ to the tuning of all the detector parameters and to set of the 
	environmental conditions on which we could act.
	The experiment sensitivity depends on factors such as the energy resolution and the live time. Therefore, we wanted to identify stable working conditions and, at the same time, 
	to improve the energy resolution by maximizing the signal-to-noise ratio.
	We performed temperature scans around the cryostat base temperature to select the one that optimized the signal and could give the designed NTD working resistance
	(a few hundreds M$\Omega$).
	
	A preliminary optimization phase occurred before the first dataset, while a second ``refined'' one was performed in between the first and the second dataset.
	In CUORE, each dataset includes one-day-long runs for a total of about one month of Physics data, and is started and ended with a calibration.
	In both datasets, the number of active channels was 984. However, during the analysis a fraction of these had to be removed for different reasons (e.\,g.\ too much noise, failure
	during one or more analysis steps, insufficient statistics collected during calibration, \dots).
	In the end, the analysis was performed on 876 channels and 935 channels, respectively.
	The average energy resolution at $Q_{\beta\beta}$, mediated over all the active channels, was $(7.7\pm 0.5)\,\keV$, with an observed improvement
	during the data collection thanks to the optimization campaign.
	The first results released by the CUORE collaboration include these two datasets, and cover the interval between May and September 2017, for a total \ce{TeO_2} exposure of 
	$86.3\,\kg\,\yr$~\cite{Alduino:2017ehq}.

\begin{figure}[t]
		\centering
		\includegraphics[width=.6\columnwidth]{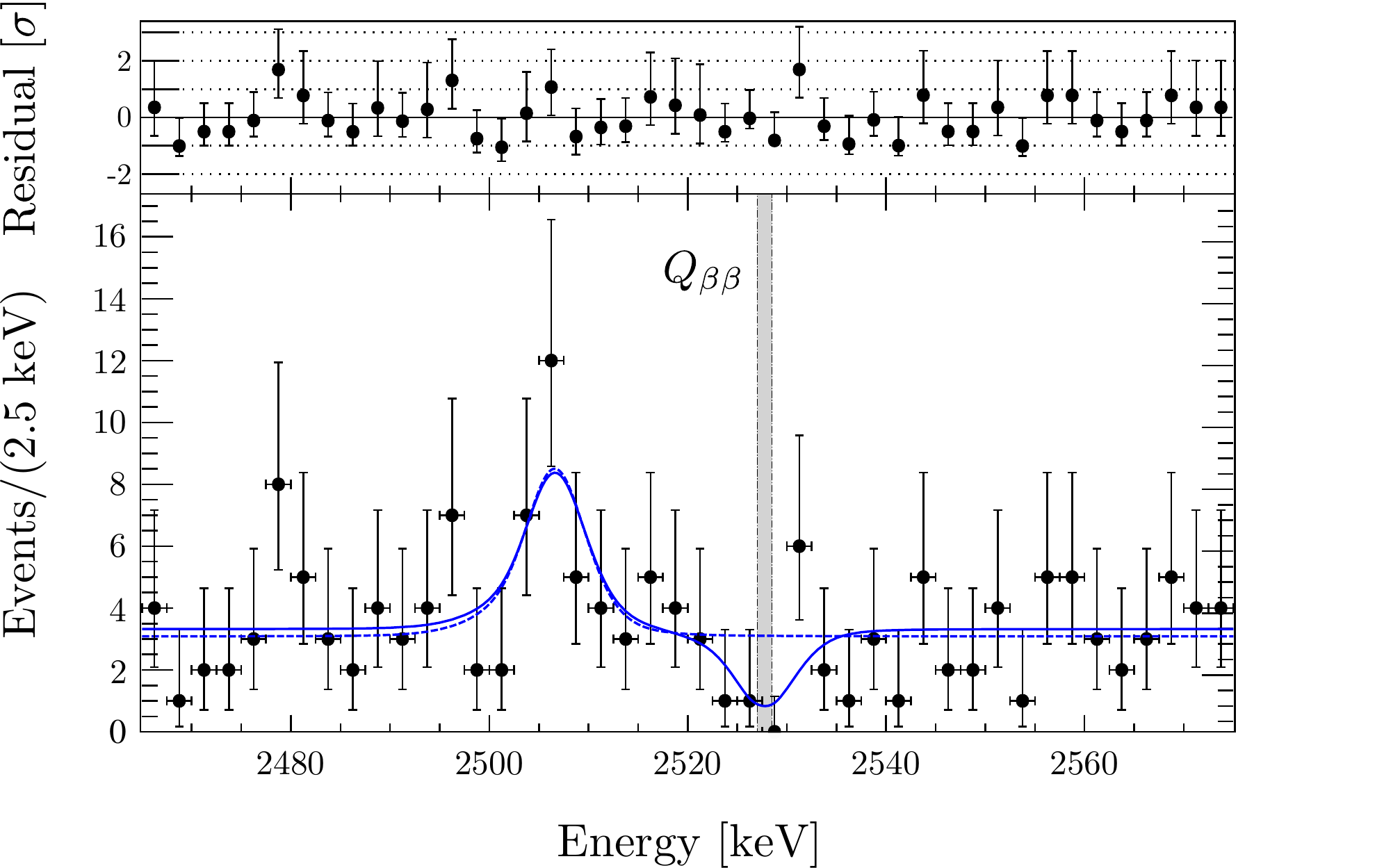}
		\caption{CUORE first data release best-fit model and normalized residuals in the ROI overlaid on the data points.
			The data are shown with Gaussian error bars. The peak at $\sim~2507$\,keV is attributed to \ce{^{60}Co}.
			The dashed line shows the continuum background component of the model. The vertical dot-dashed line indicates the position of the $Q_{\beta\beta}$ of \ce{^{130}Te}.}
		\label{fig:ROI_spectra}
	\end{figure}

	We performed a blind search for \bb. Before unblinding the actual data, we fixed the model and fitting strategy.
	We estimated the line shape parameters for each bolometer-dataset with a simultaneous, unbinned extended maximum likelihood (UEML) fit performed on each tower in the energy range 
	($2530-2720$)\,keV.
	In particular, all individual detectors were constrained to have the same decay rate, which we allowed to vary freely in the fit.
	The results is shown in Fig.~\ref{fig:ROI_spectra}, where the 155 candidate events in the Region of Interest (ROI) that passed all selection are shown, together with the UEML fit.

	We found no evidence for \bb~of \ce{^{130}Te}. Including the systematic uncertainties, we could place a lower limit on the decay half-life of $1.3\cdot 10^{25}$\,yr at 90\% C.\,L.\,.
	Combining this result with those of two earlier experiments, Cuoricino~\cite{Andreotti:2010vj} and CUORE-0~\cite{Alfonso:2015wka}, we obtained the most stringent 
	limit to date on this decay, i.\,e.\ $1.5\cdot 10^{25}$\,yr at 90\% C.\,L.\,.
	We converted the combined half-life limit as a limit on the effective Majorana neutrino mass, $\mbb$, in the framework of models that assume \bb~to mediated by light Majorana 
	neutrino exchange. We found $\mbb < (110-520)$\,meV, where the range reflects the uncertainties coming from the nuclear physics.

\section{CUORE background and \bbvv}

	In order to systematically study the CUORE radioactive contamination, we developed a background model able to describe the observed spectrum in terms of contributions from 
	contamination from the materials directly facing the detector, the whole cryogenic setup, and the environmental radioactivity.
	This detailed Monte Carlo was used over the years to guide the construction strategies of the experiment and, later, to project a background model for CUORE~\cite{Alduino:2017qet}.
	By analyzing the data from CUORE, we could ultimately test our model.

	We measured a background generally in line with or expectations: we observed an average of $(0.014\pm0.002)$\,\ckky inside the ROI.
	The contribution from $\gamma$ radiation was significantly reduced with respect to CUORE-0, and most of the $\alpha$-induced background was compatible.
	We observed an excess in the counts from \ce{^{210}Po}. Most likely, this is coming from shallow contamination in copper around the detectors, but we are still investigating it.
	Anyway, its related contribution to the ROI is estimated at level of $10^{-4}$\,\ckky.

	Thanks to our background model, we successfully reconstructed the background contribution that could be ascribed to the two-neutrino double beta decay (\bbvv) of \ce{^{130}Te}.
	Therefore, we were able to measure its half-life and we obtained $(7.9 \pm 0.1\,\mbox{\footnotesize (stat.)} \pm 0.2\,\mbox{\footnotesize (syst.)})\cdot 10^{20}\,\yr$.
	This is the world's most precise measurement on this decay.

\begin{figure}[t]
		\centering
		\includegraphics[width=.48\columnwidth]{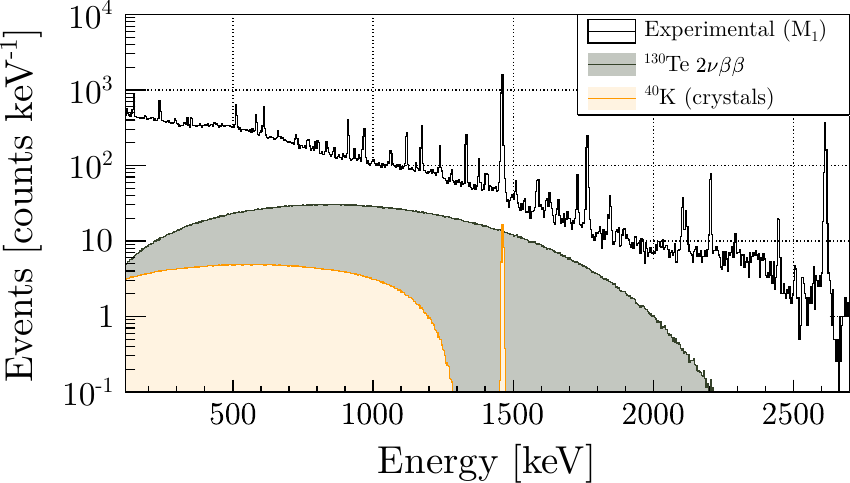} \quad
		\includegraphics[width=.48\columnwidth]{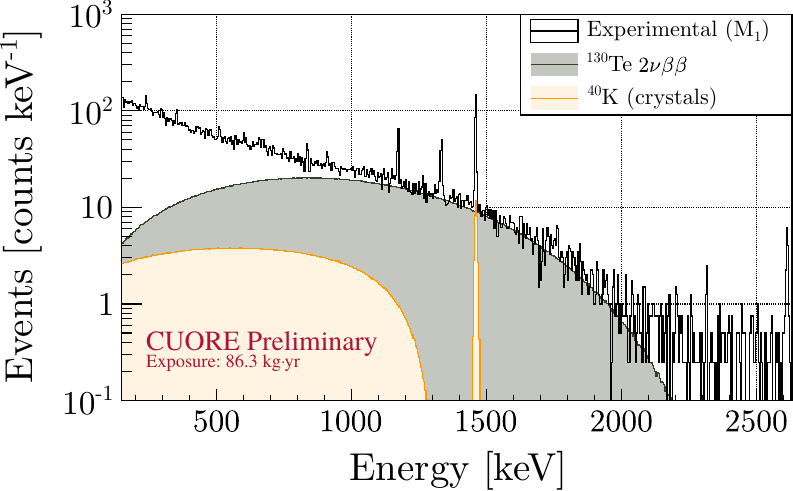}
		\caption{CUORE-0 (Left) and CUORE first data release (Right) spectra compared to the \bbvv~contribution predicted by the reference fits. 
			The \ce{^{40}K} peak from the crystal contamination (the radioactive source that has the strongest correlation with the \bbvv) is also reported.}
		\label{fig:CUORE_2nbb}
	\end{figure}

	At the same time, by comparing the contribution of the \bbvv~to the total background of CUORE with that CUORE-0 (Fig.~\ref{fig:CUORE_2nbb}), we could see that,
	while in the earlier experiment the \bbvv~spectrum accounted for $\sim 20\%$ of counts in the $(1-2)$\,MeV region, in CUORE the \bbvv~spectrum dominates
	for nearly all events in the same energy range~\cite{Alduino:2016vtd}.

\section{Outlook}

	CUORE will collect data for a total of five years of live-time. The predicted final sensitivity is $9.0 \cdot 10^{25}$\,yr at 90\%\,C.\,L.~\cite{Alduino:2017pni}.
	
	CUORE itself represents a fundamental step toward the next generation of detectors.
	Starting from the experience, the expertise, and the lessons learned while running CUORE, the CUPID project (CUORE Upgrade with Particle IDentification~\cite{Wang:2015raa}) 
	aims at developing a future bolometric \bb~experiment with sensitivity on the half-life of the order of ($10^{27}-10^{28}$)\,yr.
	Thermal detectors are expected to play a central role in the forthcoming future of the search for \bb.

\section*{Acknowledgments}

	The CUORE Collaboration thanks the directors and staff of the Laboratori Nazionali del Gran Sasso and the technical staff of our laboratories. 
	This work was supported by the Istituto Nazionale di Fisica Nucleare (INFN); the National Science Foundation under Grant Nos. NSF-PHY-0605119, NSF-PHY-0500337, NSF-PHY-0855314,
	NSF-PHY-0902171, NSF-PHY-0969852, NSF-PHY-1307204, NSF-PHY-1314881, NSF-PHY-1401832, and NSF-PHY-1404205; the Alfred P. Sloan Foundation; the University of Wisconsin Foundation; 
	and Yale University. This material is also based upon work supported by the US Department of Energy (DOE) Office of Science under Contract Nos. DE-AC02-05CH11231, DE-AC52-07NA27344, 
	and DE-SC0012654; and by the DOE Office of Science, Office of Nuclear Physics under Contract Nos. DE-FG02-08ER41551 and DE-FG03-00ER41138.  This research used resources of the National
	Energy Research Scientific Computing Center (NERSC).

\section*{References}

	\bibliography{ref}

\end{document}